\documentclass{elsart}

\usepackage{graphicx}
\usepackage{amssymb}
\usepackage[square,comma,sort&compress]{natbib}

\begin{document}

\begin{frontmatter}

\title{%
    Measurement of the radiative decay width
    \boldmath$\Gamma[\Lambda(1520)\to\Lambda\gamma]$
    with the SPHINX spectrometer
}

The SPHINX Collaboration
\author[IHEP]{Yu.~M.~Antipov},
\author[IHEP]{A.~V.~Artamonov},
\author[IHEP]{V.~A.~Batarin},
\author[IHEP]{D.~V.~Vavilov},
\author[IHEP]{V.~A.~Victorov},
\author[IHEP]{O.~V.~Eroshin},
\author[IHEP]{S.~V.~Golovkin},
\author[IHEP]{Yu.~P.~Gorin},
\author[ITEP]{V.~Z.~Kolganov},
\author[IHEP]{A.~P.~Kozhevnikov},
\author[IHEP]{A.~S.~Konstantinov},
\author[IHEP]{V.~P.~Kubarovsky},
\author[IHEP]{V.~F.~Kurshetsov},
\author[IHEP]{L.~G.~Landsberg\corauthref{cor}},
\corauth[cor]{Corresponding author.}
\ead{lgl@mx.ihep.su}
\author[IHEP]{V.~M.~Leontiev},
\author[ITEP]{G.~S.~Lomkatsi},
\author[IHEP]{V.~V.~Molchanov},
\author[IHEP]{V.~A.~Mukhin},
\author[ITEP]{A.~F.~Nilov},
\author[IHEP]{D.~I.~Patalakha},
\author[IHEP]{S.~V.~Petrenko},
\author[ITEP]{V.~T.~Smolyankin}
\address[IHEP]{Institute for High Energy Physics, Protvino, Russia}
\address[ITEP]{Institute of Theoretical and Experimental Physics, Moscow, Russia}

\begin{abstract}
The radiative decay $\Lambda(1520)\to\Lambda\gamma$ was measured directly in the study of the exclusive
diffractive-like reaction $p + N \to \Lambda(1520) K^+ + N$, 
$\Lambda(1520)\to\Lambda\gamma$ with the SPHINX spectrometer. The
values of the branching and partial width of this radiative decay were obtained:
$\mathrm{BR}[\Lambda(1520)\to\Lambda\gamma] = 
(1.02 \pm 0.21\,(\mathrm{stat}) \pm 0.15\,(\mathrm{syst}))\times 10^{-2}$
and
$\Gamma[\Lambda(1520)\to\Lambda\gamma] =
159 \pm 33\,(\mathrm{stat}) \pm 26\,(\mathrm{syst})$~keV.
\end{abstract}


\end{frontmatter}

\section{Introduction}

The study of the electromagnetic hadron decays plays an important role
in hadron spectroscopy and provides a possibility to obtain unique information
about the electromagnetic structure of 
the strongly interacting particles and about quark configurations in these hadrons. The numerous data for
radiative decays of light mesons is summarized in the reviews~\cite{Landsberg:1985fd,Zielinski:1987mg},
and for  $N^*$ and $\Delta$ baryon decays~--- in~\cite{Burkert:1992wd,Krusche:2003ik}.
At the same time the information about hyperon radiative decays is quite limited~--- see for
example~\cite{Landsberg:1996gb}.
Recent studies of the radiative hadron decays are published in~\cite{6,7,8,9,10}.

Until now the direct detection of radiative hyperon decays has been done only for two states~---
$\Sigma^0$ and $\Lambda(1520)$ hyperons. For the $\Sigma^0$ hyperon the transition

\begin{equation}
\label{1}
\Sigma^0 \to \Lambda + \gamma
\end{equation}
is the main (and practically the only) decay mode. The  $\Lambda(1520)$ hyperon may undergo decays through
various channels. These channels include, above all, the radiative transitions

\begin{equation}
\label{2}
\Lambda(1520) \to \Lambda + \gamma
\end{equation}
\begin{equation}
\label{3}
\Lambda(1520) \to \Sigma^0 + \gamma
\end{equation}

The theoretical expectations for the width of the the decay (\ref{2}) are in the wide range from 30~keV up to 215~keV,
and the ratio
$R(\Lambda/\Sigma^0) = \Gamma[\Lambda(1520)\to\Lambda\gamma] / \Gamma[\Lambda(1520)\to\Sigma^0\gamma]$
in the range from 0.26 up to 2.8 (see review~\cite{Landsberg:1996gb} and references therein).
These values are very sensitive to the SU(3) structure of the wave function of the $\Lambda(1520)$ hyperon. 
Thus the study of its radiative decays is quite important.
  
Until now these decays  were investigated in two experiments with
resonant production of  $\Lambda(1520)$ state in the low energy   $K^-p$ interactions. In the 
first experiment~\cite{Mast:1968tx} the resonant production was measured with the hydrogen 
bubble chamber in the reaction
 
\begin{equation}
\label{4}
K^- + p\to\Lambda + \mbox{(neutral particles)}
\end{equation}

Photons from radiative decays were not detected directly and were
identified by studying the missing mass (MM) spectrum with respect to
the $\Lambda$-hyperon
in the reaction~(\ref{4}) in the region (MM)$^2<0.44 m_\pi^2$.
Using this method the radiative decay~(\ref{2}) was identified and the value of the partial width 
$\Gamma[\Lambda(1520)\to\Lambda\gamma]=(134\pm 25)$~keV was determined. 
But in the estimation of the partial width the correction to the
photon spectrum influenced by the decay~(\ref{3}) was done
theoretically and was model dependent.
The value of $\Gamma[\Lambda(1520)\to\Lambda\gamma]$ in~\cite{Mast:1968tx}  after this correction 
seems to be somewhat underestimated. 

In the second experiment  \cite{Bertini:1984tx} (see also \cite{Bertini:1987tx})
with direct detection of $\Lambda(\Sigma^0)$ and $\gamma$ in the decays~(\ref{2}) and~(\ref{3})
their radiative widths were determined as $\Gamma[\Lambda(1520)\to\Lambda\gamma]=(33\pm 11)$~keV,
$\Gamma[\Lambda(1520)\to\Sigma^0\gamma]=(47\pm 17)$~keV and $R(\Lambda/\Sigma^0)=(0.71\pm 0.35)$.

Due to strong disagreement between the results of
\cite{Mast:1968tx} and \cite{Bertini:1984tx,Bertini:1987tx} for the
decay~(\ref{2}) further studies of the radiative decays
$\Lambda(1520)$ hyperon are very important. A new measurement of
$\Gamma[\Lambda(1520)\to\Lambda\gamma]$ was performed in this work in
the study of proton diffractive production reactions with the SPHINX
spectrometer in IHEP.

\section{The SPHINX spectrometer}

The SPHINX spectrometer was operating in the proton beam of the IHEP U-70 accelerator with the energy
$E_p=70$~GeV and intensity $I\simeq (2-4)\times 10^6$ $p$/spill in 1989--1999.
During that time several modifications of the detector were made  (see 
\cite{Vavilov:1994tx}, 
\cite{Bezzubov:1996tx} and  \cite{Antipov:2002tx}).
The data presented in this paper were collected with the completely upgraded latest version of the 
SPHINX spectrometer \cite{Antipov:2002tx}. The layout of the upgraded detector is
 presented in Fig.~1. The right-handed $X,Y,Z$ coordinate system of the setup had 
$Z$-axis in the direction of proton beam, vertical $Y$-axis and horizontal $X$-axis.  
The origin of  frame was in the center of magnet M. The main elements of the detector 
were as follows:
\begin{itemize}
\item[1.] The detectors of the primary proton beam~--- the scintillator counters $S_1 - S_4$ and
the scintillator hodoscopes $H_{1X,Y}, H_{2X,Y}$.
\item[2.] The targets $T_1$ (Cu; 2.64~g/cm$^2$) and $T_2$ (C; 11.3~g/cm$^2$), which were exposed 
simultaneously. The distance between the targets was 25~cm.  The
interactions of the beam in the targets were selected with the
pretrigger Str. = $S_1S_2S_3S_4(\overline{B_1B_2})$. The coincidence
$B_1B_2$ was used to reject noninteracting beam particles. The counter
system around the target region included scintillator hodoscope $H_3$
and veto counters~--- lead-scintillator sandwiches $A_1 - A_4$ (around
the targets) and $A_5 - A_8$ (in the forward direction). The holes in
the counters $A_5, A_6$ were matched with the acceptance of the
spectrometer. The information from $H_3$ and veto system could be used
to select the exclusive reactions of the diffractive production type.
\item[3.] The wide-aperture magnetic spectrometer was based on the upgraded magnet SP-40~(M) with 
a uniform magnetic field in the volume of the size $100\times 70\times 150$~cm$^3$ and 
$p_T=0.588$~GeV/$c$. The spectrometer was equipped with proportional 
chambers~(PC), drift tubes~(DT) and hodoscopes $H_{4X}$, $H_{5X}$,
$H_{6X}$, $H_{7}$, $H_{8Y}$.
The PC system
consisted of 5~X-planes and 5~Y-planes with a sensitive region of $76.8\times 64.0$~cm$^2$ and with a 2~mm wire step.
The DT system contained 18 planes of thin wall mylar tubes with a diameter of 6.25~cm. 
There were 32 tubes in each plane with wires at angles $0^{\circ};\,\pm 7.5^{\circ}$ with respect to the 
vertical $Y$-axis. The electric field in the drift tubes was described in~\cite{Antipov:1995fn}.
The sensitivity of the central tubes in the beam region was artificially reduced. The
space resolution of the DT plane was on average~$\simeq 300\,\mu$m (including the uncertainty in calibration
and in beam intensity). The hodoscopes $H_{4X},H_{5X},H_{6X},H_7$ and $H_{8Y}$ could be used 
to generate trigger signals and also in the track 
reconstruction procedure.   
\item[4.] A system of Cherenkov counters for secondary particles identification. The system included
the RICH velocity spectrometer with photo-matrix which consisted of 736 small phototubes~\cite{Vavilov:1994tx,Kozhevnikov:1999rv}
and the hodoscope-like threshold Cherenkov counter~\v{C} with 32 optically independent 
channels~\cite{Vavilov:1994tx}. The RICH 
detector was filled with SF$_6$ gas at pressure slightly above atmospheric. The threshold momenta
for $\pi/K/p$ particles in this detector were
equal to 3.5/12.4/23.6~GeV/$c$ correspondingly.
The Cherenkov detector \v{C} was filled with air at atmospheric pressure. It had 
threshold momenta of 6.0/21.3/40.1~GeV/$c$. The Cherenkov \v{C} cells were matched geometrically with the hodoscope matrix $H_7$ 
 and in principle this system could be used in the trigger for particle identification.
\item[5.] A multichannel lead glass electromagnetic calorimeter~(ECAL).
It consisted of a matrix of  $39\times 27$ cells
of the size of $5\times 5$~cm$^2$ each.
This calorimeter was previously used in the EHS experiment at
CERN~\cite{19}. One central counter was removed for the proton beam to pass through the ECAL. 
\item[6.]
A hadron calorimeter (HCAL) with a matrix of $12\times 8$
steel/scintillator total absorption counters
($5L_{\mathrm{abs}}$ thickness)~\cite{Antipov:1991}
with a size of $20\times 20$~cm$^2$ each.
\item[7.] The trigger logic system which used pretrigger Str.~=~$S_1S_2S_3S_4(\overline{B_1B_2})$ to separate beam 
interactions, hodoscope information to measure multiplicity of the secondary particles and 
information from the veto system to form different trigger signals. It was possible to work 
with 8 types of trigger signals simultaneously.
\item[8.] The front-end electronics, DAQ system and fast on-line computers. The DAQ system allowed to record 
more than 3000~events/spill.
\end{itemize}

During the runs from 1996 to 1999 on the upgraded SPHINX spectrometer
a flux of approximately 
$10^{12}$ protons has passed through 
(C/Cu) targets and more than $10^9$ trigger events were
recorded on magnetic tapes.  
These statistics are now used to study different physical processes.

\section{The \boldmath$\Lambda(1520)$ measurement and data analysis}

The study of the proton diffractive-like processes 
\begin{equation}
\label{5}
\begin{array}{rcl}
p + N(\mathrm{C}) & \to & \Lambda(1520) K^+ + N(\mathrm{C})\\
& & \kern0.36em
    \hbox{\vrule height3.0ex depth-0.6ex}
    \kern-0.37em
    \to p K^-\\[-2ex]
& & \kern0.36em
    \hbox{\vrule height5.0ex depth-0.6ex}
    \kern-0.37em
    \to \Lambda\gamma; \quad \Lambda \to p\pi^-
\end{array}
\end{equation}
at~$E_p=70\,$GeV was done with the SPHINX spectrometer while analyzing 
$[pK^-]K^+$  and $[(p\pi^-)\gamma]K^+$ systems in exclusive proton interactions with quasi-free nucleons~($N$) 
or in coherent production on C  nucleus as a whole. For this analysis the sample was obtained using the $T_{(3)}$
trigger with 3 charged particles in the final state (after $\Lambda$ decay). The well known decay channel

\begin{equation}
\label{6}
\Lambda(1520) \to p  K^-
\end{equation}
with branching ratio $\mathrm{BR}[\Lambda(1520) \to p  K^-]=(22.5\pm0.5)\%$ 
\cite{Hagiwara:2002fs} was used to study the dynamics of $\Lambda(1520)K^+$
production~(5) and to normalize the measurement of the
radiative decay branching $\mathrm{BR}[\Lambda(1520)\to\Lambda\gamma]$.

The $T_{(3)}$ trigger selected beam interactions and multiplicity of
the particles in the trigger hodoscopes which would fit events with 3
charged particles in the forward spectrometer.

\begin{equation}
\label{7}
T_{(3)} = [\mathrm{Str.} = S_1S_2S_3S_4(\overline{B_1B_2})]
H_3(0;1)H_{4X}(2;3)H_{6X}(3)
H_7(1;2;3)\bar{A}_{5-8}
\end{equation}
where $H_i(m_1;m_2)$ is the multiplicity requirement in hodoscope~$H_i$.
The selection criteria in the side hodoscope~$H_3(0;1)$
select at the same time events with quasi-free 
nucleon interactions with soft recoil protons and coherent interactions on the target carbon nuclei.

The sample selected with trigger condition $T_{(3)}$ was used to
study exclusive production of $[pK^-]K^+$ and $[\Lambda\gamma]K^+ =
[(p\pi^-)\gamma]K^+$ system in reactions

\begin{equation}
\label{8}
p+N(\mathrm{C}) \to [pK^-]K^+ + N(\mathrm{C})
\end{equation}
and
\begin{equation}
\label{9}
\begin{array}{rcl@{}l}
p + N(\mathrm{C}) & \to & [ & \Lambda\gamma]K^+ + N(\mathrm{C}) \\
& & & \kern0.36em
      \hbox{\vrule height3.0ex depth-0.6ex}
      \kern-0.37em
      \to p \pi^-
\end{array}
\end{equation}

\begin{itemize}
\item[A.] The study of the $[pK^-]K^+$ system in~(\ref{8}).

In the first step of data analysis we selected the event sample with
two positive and one negative tracks reconstructed in the magnetic
spectrometer, we also required no additional clusters in the electromagnetic
calorimeter ECAL not connected to charged tracks.
The following cuts for this event sample were applied:
\begin{itemize}
\item[1.] The primary interaction vertex $Z_1$ was fitted with the beam track in $H_{1X,Y}$, $H_{2X,Y}$
and with three secondary charged tracks in PC. The $Z$-coordinate
of the vertex $Z_1$ was required to be in the
region of the carbon target ($-555<Z_1<-525\,$~cm). The vertex resolution allowed to separate the 
interactions in carbon and copper targets and the data from copper target were not included
in the analysis in this paper).
\item[2.] The information from the RICH detector has to fit the hypothesis for the masses 
of the secondary particles. For better identification of the events the momenta of positive particles 
were required to be above 5~GeV/$c$ (threshold momentum for pions). The procedure of particle identification 
in the RICH detector was described in detail in~\cite{Kozhevnikov:1999rv}.
\item[3.] The sum of energies of the three secondary particles has to agree with primary proton energy: $65<E_p +E_{K^-} + E_{K^+}<75$~GeV.
\end{itemize} 

There was two-fold ambiguity in the identification of two positive tracks in the event as $p$ and
$K^+$. In the study of the $pK^-$ system both
mass configurations which satisfied the cuts were
included in the effective mass spectrum $M(pK^-)$. Due to good
particle identification in the RICH detector only $\sim 1\%$ of the
events allowed both mass solutions.  As was confirmed by
Monte-Carlo~(MC) simulation, the influence of these double solutions on the
determination of the number of $\Lambda(1520)\to pK^-$ events was
negligible. The measured mass spectrum $M(pK^-)$ is presented in
Fig.~2.

This spectrum is dominated by the $\Lambda(1520)$ peak. Two other peaks with $M=1.67$ and
1.8~GeV are also seen in the high mass region. In that region a dozen different hyperon 
resonances are known, and unambiguous identification of observed structures is not possible. 
The mass spectrum 
$M(pK^-)$ in Fig.~2 was fitted by the sum of smooth non-resonant background 
($P_1(M-M_\mathrm{thr})^{P_2}\cdot\exp(-P_3M-P_4M^2)$ with free parameters $P_i$ vanishing at the threshold
$M_\mathrm{thr} = m_p + m_K$) and with 3 resonance peaks.
The $\Lambda(1520)$ structure was described by a relativistic 
Breit-Wigner function smeared with the experimental resolution function
for the $pK^-$ channel. 
The total width of the $\Lambda(1520)$ was fixed to the world average value
$\Gamma = 15.6$~MeV~\cite{Hagiwara:2002fs} and a resolution of 8~MeV 
was fixed by MC
calculations. Two other peaks were described in the same way but
their masses and widths were free
parameters. The results of the fit are shown in Fig.~2 by the solid line with background shown by 
the dashed line. To avoid significant dependence of the number of resonance events on the details
of resonance parameterization in the tails of the distribution the number of events 
(here and in radiative decay) was measured in the effective 
mass region 1.40--1.65~GeV. For this region the number of  events~(\ref{6}) in reaction~(\ref{5}) 
was determined as:
$N[\Lambda(1520)\to pK^-] = 21200\pm 300$.
\item[B.] The study of the $[\Lambda\gamma]K^+ = [(p\pi^-)\gamma]K^+$ system in~(\ref{9}).

In the study of the radiative decay  $\Lambda(1520)\to\Lambda\gamma$~(\ref{2}) as well as for other 
radiative decays 
 it is very important to have a good photon veto system to reduce the 
background from different processes with additional photons (like 
$\Lambda(1520) \to \Sigma^0\pi^0,\Sigma^0\gamma$ etc.) outside and inside the acceptance of the main 
detector (see for example~\cite{Bityukov} for a detailed discussion of this method). 
In order to reduce such type of background
lead-scintillator veto counters $A_5$--$A_8$ were used as a guard
system at the trigger level in the region outside of the main acceptance
of the detector. 
The ECAL calorimeter itself was used as a part of the guard system
inside the detector acceptance in the off-line analysis,
in which the identification of one and only one photon cluster in
this calorimeter was required.

To separate the radiative decay~(\ref{2}) the sample of events
with two positive and one negative tracks in the magnetic spectrometer
and with one and only one photon cluster in ECAL
(with $E_\gamma > 1$~GeV), not associated with charged tracks,
was used to study the exclusive production of the
$[\Lambda\gamma]K^+ = [(p\pi^-)\gamma]K^+$ system in~(9).
Assuming that positive tracks are kaon and proton
and the negative track is a pion,
the reconstruction of the secondary vertex $Z_2$ was performed 
and the effective mass of $p\pi^-$  was found (with two-fold ambiguity).
Such a procedure was used to determine the $\Lambda$ track and $K^+$ track,
and using complementary information from the proton beam track
in $H_{1X,Y}$, $H_{2X,Y}$ the primary interaction vertex
$Z_1$ was reconstructed.
Additional requirements used in further analysis of this event sample
to separate exclusive production of $[\Lambda\gamma]K^+$ in~(\ref{9}) were:
\begin{itemize}
\item[1.] The primary vertex $Z_1$ had to be in the region of the main carbon target 
($-555<Z_1<-525$~cm); the secondary vertex $Z_2$ had to be along the $\Lambda$-decay path~---
before or in the 
beginning of the block of proportional chambers PC ($Z_1 + 3\sigma_{\triangle Z} < Z_2 < -280$~cm). The distance
between $Z_1$ and $Z_2$ was required to be significantly larger than the error on this measurement
($\sigma_{\triangle Z}\simeq 8$~cm and it was measured for each event).
\item[2.] The information from the RICH velocity spectrometer has to fit assumptions about the masses of secondary particles (see discussion above in section~A). As
before, for better identification of the particles the momenta  of positive tracks were
required to be $>5$~GeV/$c$.
\item[3.]
The effective mass of the $p\pi^-$ system was required
to be in the $\Lambda$ mass region $1.106 < M(p\pi^-) < 1.126$~GeV.
\item[4.] The sum of the energies of secondary particles had to agree with the energy of the
proton beam. To reduce the background from reactions with lost photons more stringent cuts
 than in  sample A were used:
$68 < E_\Lambda + E_{K^+} + E_\gamma <73$~GeV.
\end{itemize}

Due to two positive particles in the final state the two-fold ambiguity in proton-kaon identification took place.
For further analysis both solutions which satisfied all the requirements were used.
But due to strong constraints on the $\Lambda$ hyperon vertex and good particle identification in the 
RICH detector the number of double solutions was negligible. The effective mass spectrum $M(p\pi^-)$ in~(9)
for the sample B after all cuts (except for the cut on the effective mass itself) is presented in Fig.~3.
It is dominated by the $\Lambda$ peak with a very small background. The effective mass spectrum 
$M(\Lambda\gamma)$ in the sample B is shown in Fig.~4.

In the $\Lambda\gamma$ spectrum a clear signal with a mass of the $\Lambda(1520)$ hyperon is seen as well as
additional structures below and above the $\Lambda(1520)$ peak. The 
background in the intermediate region can be explained
partly by some 
resonant decays with lost photon and partly by non-resonant background with soft fake ``photons''
(the noise in the ECAL due to charged particle interactions in the spectrometer).

To study the $M(\Lambda\gamma)$ spectrum in Fig.~4 and to develop the normalization procedure to
measure $\mathrm{BR}[\Lambda(1520)\to\Lambda\gamma]$ we performed a very detailed MC simulation
of the SPHINX spectrometer based on the package GEANT-3.21. This simulation procedure described the 
SPHINX geometry and the materials, efficiency and resolutions of all detectors, including the
reduced efficiency of PC and DT in the beam region. The responses from all detectors were simulated including the
digitization. All secondary tracks and backflashes were traced and accounted.
In the final analysis the full simulation procedure was used with the 
same reconstruction programs which analyzed the raw data. Kinematic characteristics of  
$\Lambda(1520)K^+$ production in reaction~(\ref{5}) (distributions for momentum transfer,  
effective mass and angles in the Gottfried-Jackson frame) were 
studied with the decay~(\ref{6}) and were simulated to describe
 proper distributions of the experimental data. 

It was shown in the analysis of the mass spectrum in Fig.~4 that structure in the region of 1.35~GeV 
corresponded to decays $\Sigma(1385)^0\to\Lambda\pi^0$ and $\Lambda(1405)\to\Sigma^0\pi^0$ with one 
or two lost photons. The structure in the region 1.7~GeV corresponded to decays of higher mass hyperon 
resonances. The smooth intermediate background is connected with the decays $\Lambda(1520)\to\Sigma^0\pi^0$,
$\Lambda(1520)\to\Sigma^0\gamma$ with lost photons and with decays $\Sigma(1385)^0\to\Lambda\gamma$,
 $\Lambda(1405)\to\Lambda\gamma$.
We demonstrate for example in Fig.~5 the MC simulation of the $\Lambda\gamma$ spectrum from radiative decay
$\Lambda(1520)\to\Lambda\gamma$ and background processes $\Lambda(1520)\to\Sigma^0\pi^0$ and
$\Lambda(1520)\to\Sigma^0\gamma$    with two and one lost photons. In spite of some uncertainties
connected with the unknown branching ratio of $\Lambda(1520)\to\Sigma^0\gamma$ it is possible to conclude
that these backgrounds from decays with lost photons are not very significant and are shifted to the
region of smaller masses. But certainly all these background effects must be taken into consideration 
in the final analysis of the $M(\Lambda\gamma)$ spectrum in Fig.~4.

To determine the number of events of the radiative decay $\Lambda(1520)\to\Lambda\gamma$ the $M(\Lambda\gamma)$
spectrum in Fig.~4 was fitted by the sum of the $\Lambda(1520)\to\Lambda\gamma$ resonance and background. The resonance
$\Lambda(1520)$ was presented in the relativistic Breit-Wigner form with table width 
$\Gamma = 15.6$~MeV smeared with the experimental resolution $\sigma = 26$~MeV (from MC calculations).
The mass of the $\Lambda(1520)$ was considered as free parameter. Any possible influence of decay~(\ref{2}) dynamics
(with allowed orbital momentum values $L=0$ and $L=2$) on the fit was negligible~($\lesssim 2\%$). 
The background in this spectrum
was described as a sum of a smooth function $P_1\cdot \exp(-P_2M - P_3M^2)$ and three normal distributions 
corresponding to the peaks with $M=1.35$~GeV and 1.7~GeV and to a third effective peak
which took into consideration background processes $\Lambda(1520)\to\Sigma^0\pi^0$,
$\Lambda(1520)\to\Sigma^0\gamma$, $\Sigma(1385)^0\to\Lambda\gamma$, $\Lambda(1405)\to\Lambda\gamma$.
Parameters $P_i$ in the smooth background and all parameters of the 3 normal distributions for the 3 
additional peaks were free in the fit. The result of the fit is presented
in Fig.~4 as a solid line. The yield of $\Lambda(1520)\to\Lambda\gamma$  radiative
decay and the resonance backgrounds with lost photons are shown in Fig.~4 as dashed curves.
\end{itemize}

As a result of the fit we measured the number of radiative decays~(\ref{2}) in reaction~(\ref{5}) in the mass region 
1.40--1.65~GeV (the same as in 
$\Lambda(1520)\to pK^-$): $N[\Lambda(1520)\to\Lambda\gamma] = 290\pm 60$ events. From the MC 
simulation the ratio of efficiencies $\varepsilon = \varepsilon_2[\Lambda(1520)\to\Lambda\gamma]
/\varepsilon_6[\Lambda(1520)\to pK^-] = 0.47$ was calculated. From these data, the ratio of the decay branching
ratios was obtained: 
\begin{eqnarray}
\label{10}
\frac{\mathrm{BR}[\Lambda(1520)\to\Lambda\gamma]}
     {\mathrm{BR}[\Lambda(1520)\to pK^-]} & = & \frac{N[\Lambda(1520)\to\Lambda\gamma]/\mathrm{BR}
      (\Lambda\to p\pi^-)}
     {N[\Lambda(1520)\to pK^-]\cdot \varepsilon} = \nonumber \\  
 = \frac{(290\pm 0.60)/0.639}{(21200\pm 30)\cdot 0.47} & = &   (4.55\pm0.94)\times 10^{-2}
\end{eqnarray}

From~(\ref{10}) and from table values for branching ratio and width of 
$\Lambda(1520)\to pK^-$~\cite{Hagiwara:2002fs} the final
values for radiative decay~(\ref{2}) were determined:
\begin{equation}
\label{11}
\left. \begin{array}{l}
\mathrm{BR}[\Lambda(1520)\to\Lambda\gamma] =
[1.02 \pm 0.21\,(\mathrm{stat}) \pm 0.15\,(\mathrm{syst})]\times 10^{-2} \\
\Gamma[\Lambda(1520)\to\Lambda\gamma] =
[159 \pm 33\,(\mathrm{stat}) \pm 26\,(\mathrm{syst})]\,\mbox{keV}
\end{array}\right\}
\end{equation}

Many systematic errors connected with the relative measurement of
$\Lambda(1520)\to\Lambda\gamma$ and $\Lambda(1520)\to pK^-$ decays
canceled out (trigger efficiency, reconstruction efficiency,
accidental losses etc.). Two remaining main sources of systematics are:
\newline
a)~The photon efficiency systematics.
\newline
We estimate this value by studying many processes with one photon,
two photons and many photons in the final state. Many of these
processes were discussed in our previous publications (see, for
example, \cite{Vavilov:1994tx,Bezzubov:1996tx,Antipov:2002tx}
and references therein) and in a new paper~\cite{23}.
By comparing results of these studies with our GEANT MC
simulations we estimate the systematic error in photon efficiency
in the decay~(\ref{2}) as~$7\%$.
\newline
b)~The fitting systematics for $M(\Lambda\gamma)$ in Fig.~\ref{fig-mc-bgr}.
\newline
We performed many variations of this fit by changing the smooth background
function and excluding different background structures in the fit
or changing their parameters. The variation of the number of the
$\Lambda(1520)\to\Lambda\gamma$ events in the signal peak has been
used to estimate the systematic uncertainties of the fitting procedure.
This uncertainty is estimated as~$13\%$.

Thus our total systematic uncertainty in the branching ratio
$\Lambda(1520)\to\Lambda\gamma$ is~$15\%$.
The systematic uncertainty in the partial width of this decay
is slightly higher because of the uncertainty
in the world average value for the total $\Lambda(1520)$ width.
These numbers are included in expression~(\ref{11}).

It is possible to reduce significantly the background level in the study of $M(\Lambda\gamma)$
mass spectrum if we use only events without interactions of charged particles in the ECAL calorimeter. Cleaner
conditions in ECAL for this event sample allowed a reduction of the energy threshold of a photon
cluster $E_{\gamma} > 0.5$~GeV and thus to reduce background. The $M(\Lambda\gamma)$ spectrum with this 
additional requirement is presented in Fig.~6. In spite of smaller statistics in this sample the peak of
the $\Lambda(1520)\to\Lambda\gamma$ decay is observed more clearly and background is reduced 
significantly. But we use this result only for demonstrative purposes,
to stress qualitatively
the observation of the $\Lambda(1520)\to\Lambda\gamma$ radiative decay.
We do not use this result for  quantitative measurements
because of significantly reduced statistics and some
additional systematic uncertainties. This figure
demonstrates the reduction of the lost photon background
after reduction of the photon thresholds in ECAL. We are working
now on other algorithms of ECAL analysis hoping that
in future studies it would be possible to reduce this threshold
without such large losses in statistics. If we are successful
it gives us the possibility to reduce background and systematics
and to obtain the results for the decay~(\ref{3}).

\section{Conclusion}

In the SPHINX spectrometer the radiative decay
$\Lambda(1520)\to\Lambda\gamma$ was studied in exclusive proton
reaction~(\ref{5}) and its partial width and branching ratio were measured.
The result of this measurement in existing precision level agrees with
previous measurement~\cite{Mast:1968tx}, but is in strong
contradiction with~\cite{Bertini:1984tx,Bertini:1987tx}. We plan to
study in more detail the systematic uncertainties of our measurement
and to try to reduce them.  We also plan to study the possibility to
measure the radiative decay $\Lambda(1520)\to\Sigma^0\gamma$ in the SPHINX
experiment.

\section*{Acknowledgements}
We would like to thank Prof. B. Povh and Prof. R. Bertini for the discussion of the results of their 
experiments \cite{Bertini:1984tx,Bertini:1987tx}. This work was partly supported by Russian
Foundation for Basic Researches 
(grants 99-02-18252 and~02-02-16086).

\newpage

\begin{figure}[h]
\includegraphics[width=\textwidth]{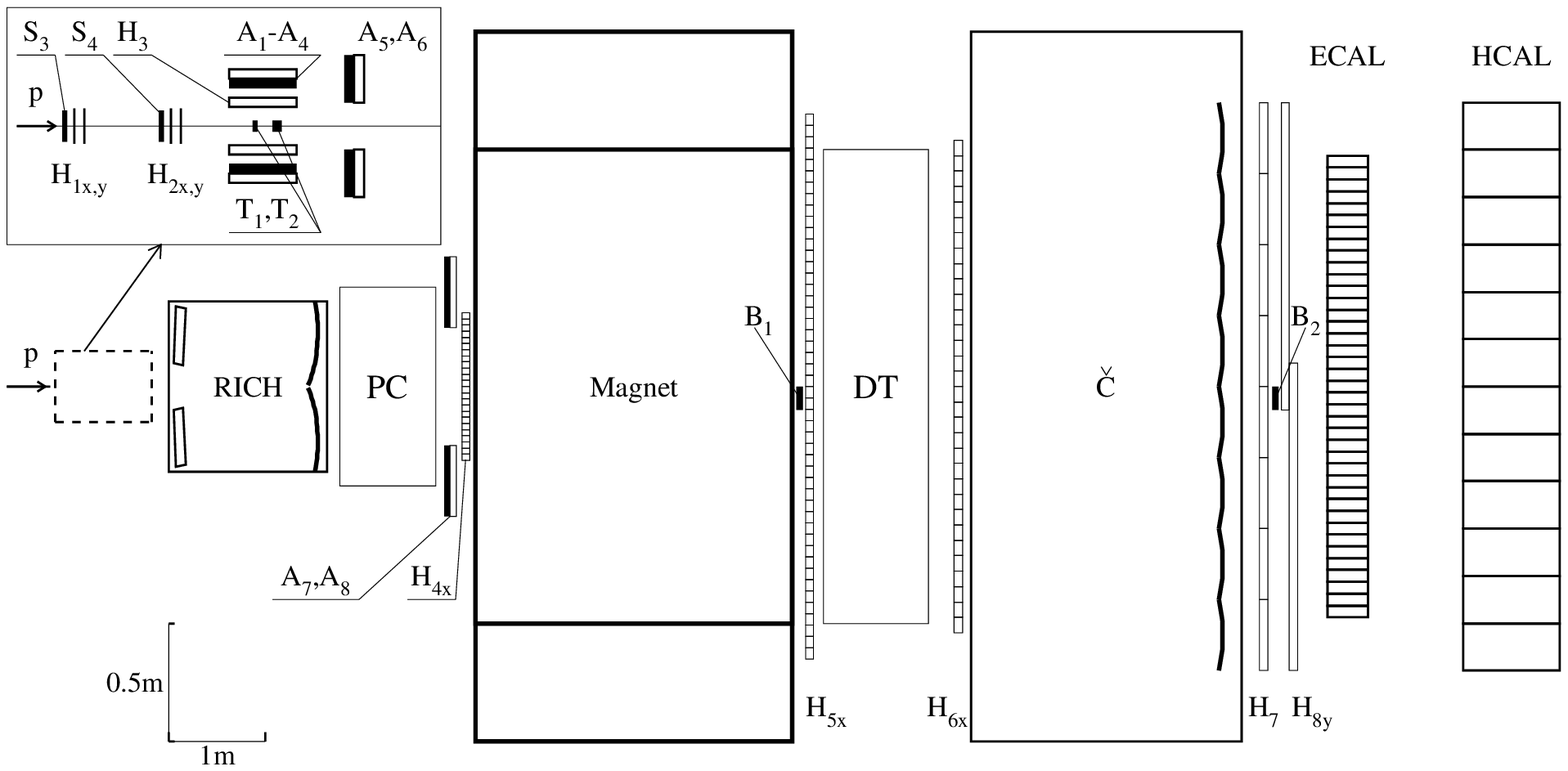}
\caption{%
Layout of the SPHINX spectrometer:
$\mathrm{S}_1$--$\mathrm{S}_4$~---
beam scintillator counters
(the very upstream counters $\mathrm{S}_1$ and $\mathrm{S}_2$
are not shown in this Fig.);
$\mathrm{T}_1$,~$\mathrm{T}_2$~---
copper and carbon targets;
$\mathrm{A}_1$--$\mathrm{A}_8$~---
lead-scintillator veto counters;
$\mathrm{B}_1$--$\mathrm{B}_2$~---
veto counters to tag non-interacting primary beam particles;
$\mathrm{H}_{1X,Y}$, $\mathrm{H}_{2X,Y}$~--- beam hodoscopes;
$\mathrm{H}_3$~--- side hodoscope around the target;
$\mathrm{H}_{4X}$, $\mathrm{H}_{5X}$, $\mathrm{H}_{6X}$, $\mathrm{H}_{7}$, $\mathrm{H}_{8Y}$~---
hodoscopes in magnetic spectrometer;
PC~--- proportional chambers and
DT~--- drift tubes of magnetic spectrometer;
RICH~--- velocity spectrometer to register the rings of Cherenkov radiation;
\v{C}~--- multichannel threshold Cherenkov hodoscope;
ECAL~--- electromagnetic calorimeter;
HCAL~--- hadron calorimeter (for details see text).
}
\label{fig-sphinx21}
\end{figure}

\begin{figure}[h]
\includegraphics{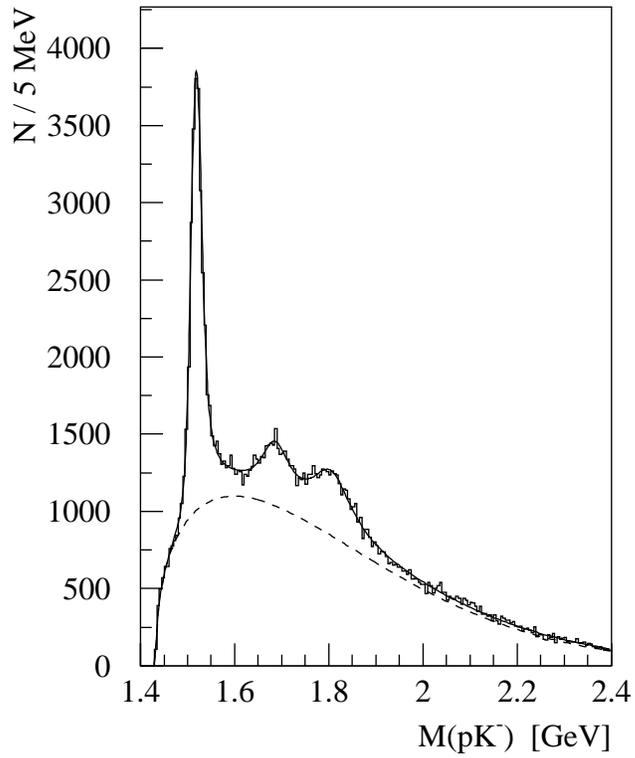}
\caption{%
Effective mass spectrum  $M(pK^-)$ in the reaction  
$p + N(\mathrm{C}) \to pK^-K^+ + N(\mathrm{C})$
(8). In this spectrum the peak of
 $\Lambda(1520)\to pK^-$ dominates.
The fitted spectrum corresponds to the solid line (see text).
Non-resonant background is shown by the dashed line.}
\label{fig-mass-p+pi-}
\end{figure}

\begin{figure}[h]
\includegraphics{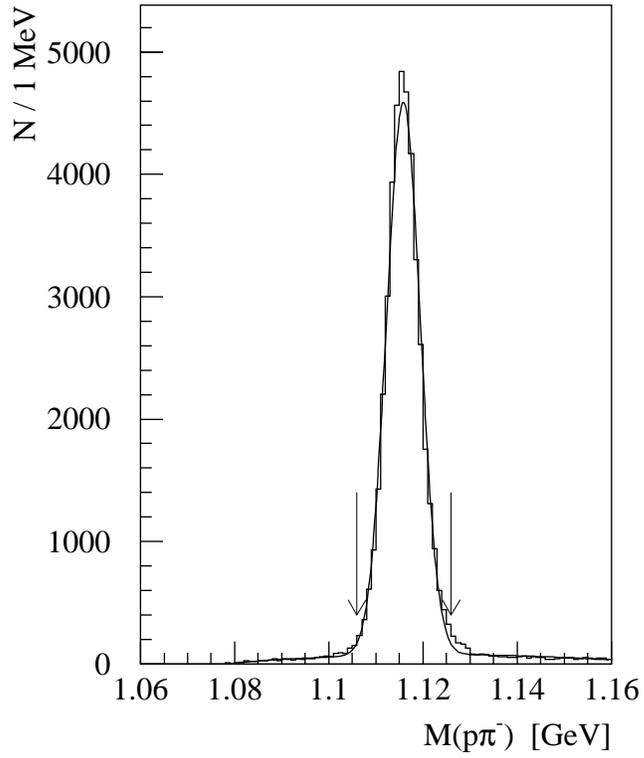}
\caption{%
The mass spectrum $M(p\pi^-)$ in reaction
 $p + N(\mathrm{C}) \to [(p\pi^-)\gamma] K^+ + N(\mathrm{C})$~(9) (see text).
This spectrum is dominated by the $\Lambda\to p \pi^-$ peak.
}
\label{fig-mass-lambda+gamma}
\end{figure}

\begin{figure}[h]
\includegraphics{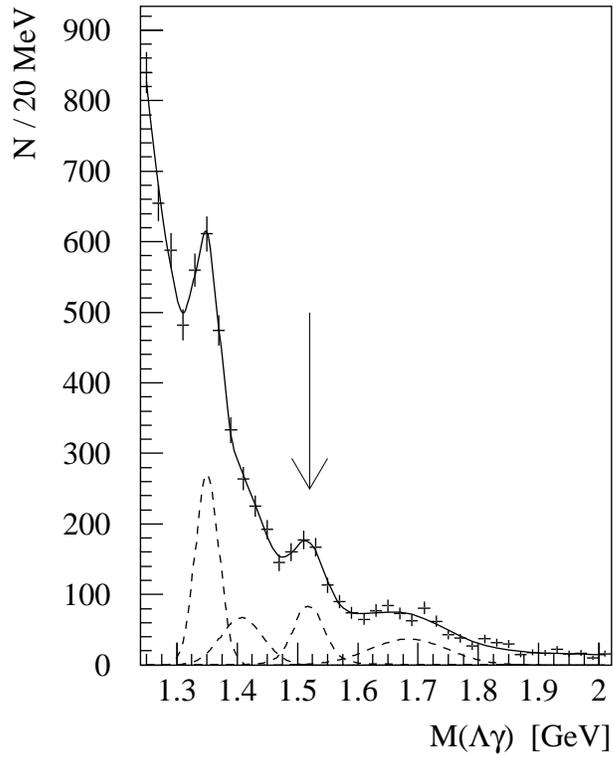}
\caption{%
Effective mass spectrum  $M(\Lambda\gamma)$ in reaction~(9). The fit of this spectrum
is described in the text and is shown by the solid line. The dashed line shows the
$\Lambda(1520)\to\Lambda\gamma$ peak (see the arrow)
and background resonant contributions.
}
\label{fig-mc-bgr}
\end{figure}

\begin{figure}[h]
\includegraphics{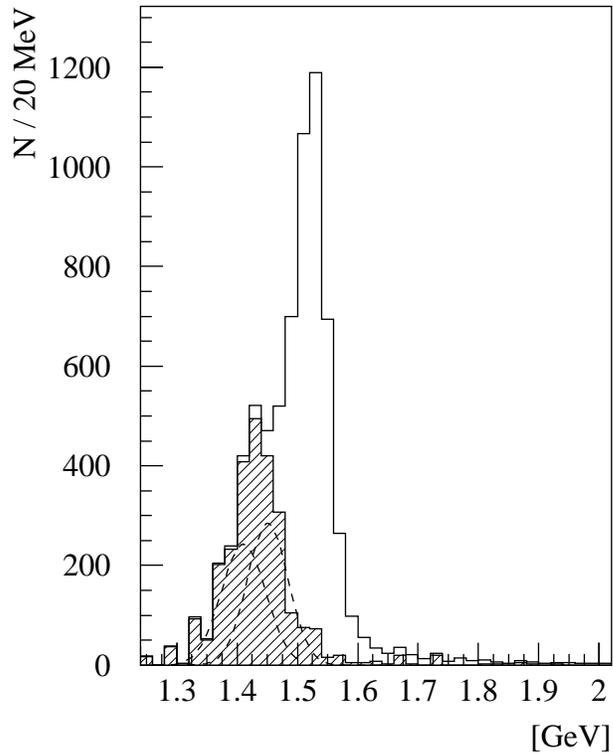}
\caption{%
Monte-Carlo study of the effective mass spectra, described in the text.
The main peak corresponds to the decay $\Lambda(1520)\to\Lambda\gamma$ (for
$\mathrm{BR}[\Lambda(1520)\to\Lambda\gamma] = 1\%$). The peaks in the region of smaller masses are
from processes with  one and two lost photons $\Lambda(1520)\to\Sigma^0\gamma$,
 (assuming the $\mathrm{BR}[\Lambda(1520)\to\Sigma^0\gamma] = 1\%$)
and $\Lambda(1520)\to\Sigma^0\pi^0$ with $\mathrm{BR}[\Lambda(1520)\to\Sigma^0\pi^0] = 14\%$~---
see~\cite{Hagiwara:2002fs}.
}
\end{figure}

\begin{figure}[h]
\includegraphics{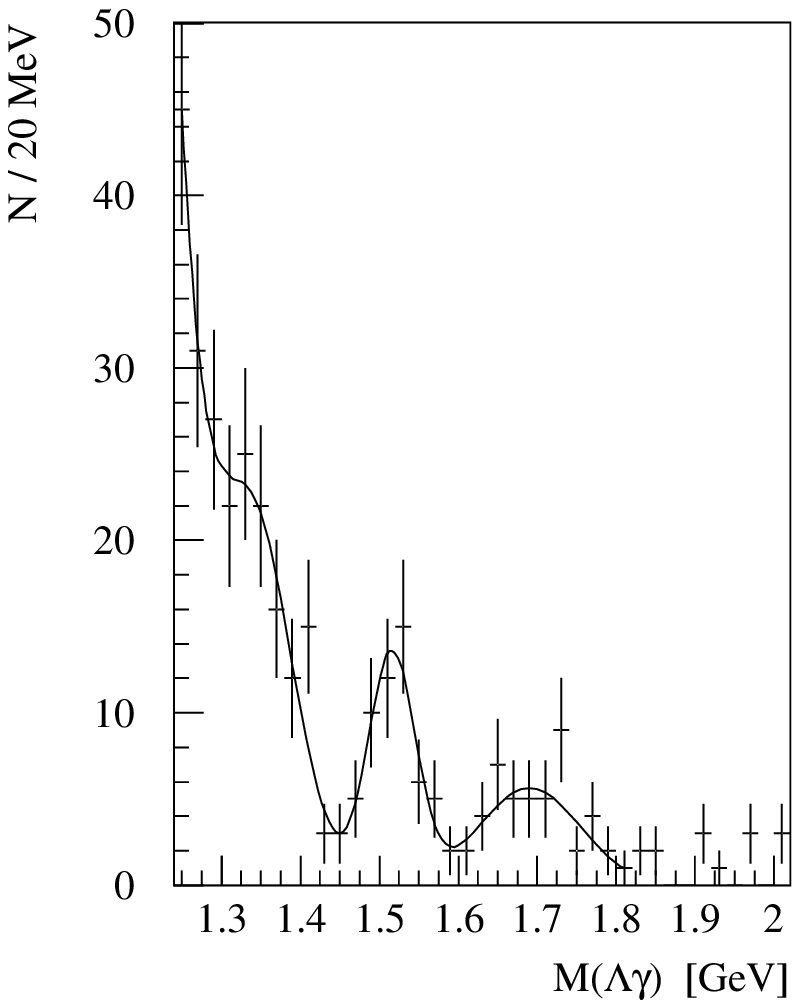}
\caption{%
Effective mass spectrum  $M(\Lambda\gamma)$ in the reaction  
$p+N(\mathrm{C})\to[\Lambda\gamma]K^+
+N(\mathrm{C})$
(9) obtained for more stringent criteria for suppression of background: without interactions of
charged particles in ECAL and with a reduced threshold for one and only one photon
cluster in this spectrometer~($E_{\gamma}>0.5$~GeV). This spectrum is used only for good
 qualitative demonstration of the observation of the radiative decay
  $\Lambda(1520)\to\Lambda\gamma$.}
\end{figure}


\begin{thebibliography}{00}
\bibitem{Landsberg:1985fd} L. G. Landsberg, Phys. Rep. {\bf 128}, 301 (1985); Yad. Fiz. {\bf 63},
        3 (2000),  [Engl. Transl. Phys. Atom. Nucl. {\bf 63}. 5 (2000)].
\bibitem{Zielinski:1987mg} M. Zielinski, Acta Phys. Polon. B {\bf 18}, 455 (1987).
\bibitem{Burkert:1992wd} D. Burkert, Int. J. Mod. Phys. E {\bf 1}, 421 (1992).
\bibitem{Krusche:2003ik} B. Krusche and S. Schadmand, Prog. Part. Nucl. Phys. {\bf 51}, 399 (2003),
         nucl-ex/0306023.
\bibitem{Landsberg:1996gb} L. G. Landsberg, Yad. Fiz. {\bf 59}, 2161 (1996),
        [Engl. Transl. Phys. Atom. Nucl. 59, 2080 (1996)].
\bibitem{6} V. V. Molchanov {\it et al.}, (SELEX Collab.), Phys. Lett. B {\bf 521}, 171 (2001).
\bibitem{7} V. V. Molchanov {\it et al.}, (SELEX Collab.), Phys. Lett. B {\bf 590}, 161 (2004)
           [hep-ex/0402026].
\bibitem{8} A. Aloisio {\it et al.}, (KLOE Collab.), Proceedings of Hadron Spectroscops (Hadron-2001) Protvino,
            Russia, 25 August - 1 September 2001, p. 711. p. 716.
\bibitem{9} M. N. Achasov {\it et al.}, (SND Collab.), Proceedings of Hadron Spectroscops (Hadron-2001) 
            p. 30, p. 72, Protvino, Russia, 25 August - 1 September 2001.
\bibitem{10} D. Alde {\it et al.}, (GAMS Collab.), Phys. Lett. B {\bf 340}, 122 (1994).
\bibitem{Mast:1968tx} T. S. Mast {\it et al.}, Phys. Rev. Lett. {\bf 21}, 1715 (1968).
\bibitem{Bertini:1984tx} R. Bertini {\it et al.}, Contribution NM18 of Heidelberg-Saclay-Strasbourg 
       Collab. at PANIC-84 (Particles and Nuclei 10th International Conference), Heidelberg, 1984.
\bibitem{Bertini:1987tx} R. Bertini, Nucl. Phys. B {\bf 279}, 49 (1987).
\bibitem{Vavilov:1994tx} D. V. Vavilov {\it et al.}, (SPHINX Collab.), Yad. Fiz. {\bf 57}, 241 (1994),
     [Engl. Transl. Phys. Atom. Nucl. {\bf 57}, 227 (1994)].
\bibitem{Bezzubov:1996tx} V. A. Bezzubov {\it et al.}, (SPHINX Collab.), Yad. Fiz. {\bf 59}, 2199 (1996),
     [Engl. Transl. Phys. Atom. Nucl. {\bf 59}, 2117 (1996)].
\bibitem{Antipov:2002tx} Yu. M. Antipov {\it et al.}, (the SPHINX Collab.), Yad. Fiz. {\bf 65}, 2131 (2002),
                        [Engl. Transl. Phys. Atom. Nucl. {\bf 65}, 2070 (2002)].
\bibitem{Antipov:1995fn} Yu. M. Antipov {\it et al.}, Nucl. Phys. Proc. Suppl. {\bf 44}, 206 (1995).
\bibitem{Kozhevnikov:1999rv} A. P. Kozhevnikov {\it et al.}, Nucl. Instrum. Meth. A {\bf 433}, 164 (1999).
\bibitem{19} B. Powell {\it et al.}, Nucl. Instr. Methods {\bf 198}, 217 (1982).
\bibitem{Antipov:1991} Yu. M. Antipov {\it et al.}, PTE {\bf 1}, 45 (1991).
\bibitem{Hagiwara:2002fs} K. Hagiwara {\it et al.}, (Particle Data Group), Phys. Rev. D {\bf 66}, 
         010001 (2002).
\bibitem{Bityukov} S. I. Bityukov {\it et al.}, (Lepton Collab), Yad. Fiz. {\bf 47}, 1258 (1988),
           [Engl. Transl. Sov. J. Nucl. Phys. {\bf 47}, 800 (1988)].
\bibitem{23}
Yu. M. Antipov et al. (The SPHINX Collaboration). hep-ex/0407026
(to be published in Eur. Phys.~J.~A).
\end{thebibliography}
\end{document}